\documentclass[aip,pop,reprint]{revtex4-1}
\usepackage{graphicx}
\usepackage{ulem}
\usepackage{amssymb}
\usepackage{amsmath}
\usepackage{todonotes}
\usepackage{wasysym}
\usepackage{relsize}
\usepackage{bm}
\begin{document}


 

\title{Shielding effects in random large area field emitters, the field enhancement factor distribution and current calculation}

\author{Debabrata Biswas}
\affiliation{
Bhabha Atomic Research Centre,
Mumbai 400 085, INDIA}
\affiliation{
Homi Bhabha National Institute, Mumbai 400 094, INDIA}
\author{Rashbihari Rudra }
\affiliation{
Bhabha Atomic Research Centre,
Mumbai 400 085, INDIA}


\begin{abstract}
  A finite-size uniform random distribution of vertically aligned field emitters on a planar surface is
  studied under the assumption that the asymptotic field is uniform and parallel to the
  emitter axis. A formula for field enhancement factor is first derived for a 2-emitter system and
  this is then generalized for $N$-emitters placed arbitrarily (line, array or random).
  It is found that geometric effects dominate the shielding of field lines.
  The distribution of field enhancement factor for a uniform random distribution of emitter locations
  is found to be closely approximated by an extreme value (Gumbel-minimum) distribution when the mean separation
  is greater than the emitter height but is better approximated by a Gaussian for mean separations close to the
  emitter height. It is shown that these distributions 
  can be used to accurately predict the current emitted from a large area field emitter.
\end{abstract}






\maketitle

\newcommand{\be}{\begin{equation}}
\newcommand{\ee}{\end{equation}}
\newcommand{\bea}{\begin{eqnarray}}
\newcommand{\eea}{\end{eqnarray}}
\newcommand{\Tbar}{{\bar{T}}}
\newcommand{\En}{{\cal E}}
\newcommand{\K}{{\cal K}}
\newcommand{\GC}{{\cal \tt G}}
\newcommand{\Lop}{{\cal L}}
\newcommand{\DB}[1]{\marginpar{\footnotesize DB: #1}}
\newcommand{\q}{\vec{q}}
\newcommand{\kt}{\tilde{k}}
\newcommand{\Lopn}{\tilde{\Lop}}
\newcommand{\noi}{\noindent}
\newcommand{\ovn}{\bar{n}}
\newcommand{\ovx}{\bar{x}}
\newcommand{\ovE}{\bar{E}}
\newcommand{\ovV}{\bar{V}}
\newcommand{\ovU}{\bar{U}}
\newcommand{\ovJ}{\bar{J}}
\newcommand{\calE}{{\cal E}}
\newcommand{\ovphi}{\bar{\phi}}
\newcommand{\zt}{\tilde{z}}
\newcommand{\rt}{\tilde{\rho}}
\newcommand{\tth}{\tilde{\theta}}
\newcommand{\nuv}{{\rm v}}
\newcommand{\ck}{{\cal K}}
\newcommand{\cc}{{\cal C}}
\newcommand{\ca}{{\cal A}}
\newcommand{\cb}{{\cal B}}
\newcommand{\cg}{{\cal G}}
\newcommand{\ce}{{\cal E}}
\newcommand{\fn}{{\small {\rm  FN}}}
\newcommand\norm[1]{\left\lVert#1\right\rVert}



\section{Introduction}
\label{sec:intro}

The in-principle advantages of using field emission cathodes over thermionic ones are manifold.
Issues associated with high temperature operation and temporal response in thermionic cathodes clearly
indicate that next-generation high performance electron emission systems for use in vacuum
devices must be based on field emission \cite{teo,parmee}. The strides achieved in the past decades in our ability
to pattern arrays of pointed emitters \cite{spindt68,spindt76,spindt91} and the discovery of
carbon nanotubes (CNT) as a suitable material \cite{deHeer,cole2014} 
for stable operation, have led to vigorous research activity in this direction. These efforts are
supported by theoretical studies on large area field emitters (LAFE) \cite{zhbanov,forbes2012,read_bowring,
harris15,forbes2016,jensen_ency,jap2016} and corrections for nano-tipped emitters \cite{jensen_image,db_imag,db_ext} to the planar Fowler-Nordheim (FN) formula for current density \cite{FN,murphy,forbes,forbes_deane}.
Despite this progress, there are several challenges in our ability to predict theoretically, the current emitted by
a single emitter or a cluster of emitters placed randomly or in an array.

The Fowler-Nordheim formalism continues to remain relevant despite the vast change in experimental
field-emission setups. This is surprising considering that it is based on a planar model
for metallic emitters. Field emitters of today are highly curved leading to large
local electric fields at the apex due to the phenomenon
known as field-enhancement. Thus, moderate asymptotic fields ($\sim$ V/$\mu$m) can lead to
local fields as large as 5-10 V/nm, at which appreciable field emission can occur.
The transition from planar to curved emitters in field-emission theory is generally made
using a local apex field enhancement factor (AFEF) $\gamma_a$ \cite{forbes2003,db_fef}.
It does not change the shape of
traditional FN-plots and allows one to extract the parameter $\gamma_a$.
However, theoretical estimates of the emission current require knowledge about the AFEF
and efforts in this direction depend either on analytically tractable models such as the
hemisphere or hemiellipsoid on a plane in the presence of an asymptotic field $E_0$
that is uniform and parallel to the emitter axis or rely on finite-element codes
for particular emitter shapes and diode configuration. Using a different approach,
a recent study using the line charge model (LCM), generalizes the known result for the hemiellipsoid  
and expresses the apex field enhancement factor as 

\be
\gamma_a = \frac{2h/R_a}{\alpha_1 \ln(4h/R_a) - \alpha_2}  \label{eq:fef0}
\ee

\noi
where  $h$ is the height of the
emitter, $R_a$ the apex radius of curvature and $\alpha_1,\alpha_2$
depend on the details of the line charge and hence the emitter shape. It was also found numerically that
the field enhancement factor is equally well described by the simpler form

\be
\gamma_a = \frac{2h/R_a}{\ln(4h/R_a) - \alpha_0}  \label{eq:fef}
\ee

\noi
where $\alpha_0$ was found to depend on the emitter base (e.g. cone, ellipsoid, cylinder).
Single emitter predictions for emitter current can thus be made under the condition
that the image charges at the anode can be neglected (large anode-cathode separation)
and the work function and band-structure variations on the active emission surface is
negligible \cite{db2018_pop1,db_fef}.

While single emitter setups are important in their own right, an efficient and bright
electron source requires a large area field emitter comprising of numerous emission
tips (such as CNTs) placed in an array or even randomly. From a theoretical perspective,
despite all simplifying assumptions, there is the added complication of each emitting
site in a finite-sized patch, having a different enhancement factor due to the
process of shielding. Emitters in close proximity ``shield'' an emitter apex
thereby lowering the enhancement factor from its un-shielded value. While attempts
have been made to understand the shielding process using models such as the floating-sphere
on emitter plane potential as well as numerically, a
quick and accurate prediction of the apex field enhancement factors in a LAFE based on the
proximity of the other emitters, or even an estimate of the average enhancement factor for a given mean
separation, is not available.

We shall deal here with a cluster of emitters, all having the same height and apex radius of curvature but
placed randomly following a uniform distribution on a rectangular patch. Our methods allow us to deal
with arrays as well. Our interest is twofold. First, we shall try to understand the process
of shielding and try to arrive at a simple unifying picture. Next, we shall probe the existence of a universal
field enhancement factor distribution \cite{groening2001,cole2014,groening2015} when the
emitters are placed randomly. This is then used to
find the net emission current from a random LAFE.
Our approach here is a generalization of the method adopted recently \cite{db_fef} to arrive at Eq.~\ref{eq:fef0}.
We shall first introduce the line charge model and consider the case of two emitters. The result can
then easily be generalized to $N$-emitters.

\section{The 2-Emitter case - line charge model}

Consider two emitters of height $h$, apex radius of curvature $R_a$, separated by a distance $\rho_{12}$,
placed on a grounded metallic plane and aligned along an asymptotic (away from the emitter tips) electrostatic field
$-E_0 \hat{z}$. If we assume the first emitter to be centred at the origin such that its apex has co-ordinates
$(\rho,z) = (0,h)$ while the apex of the second emitter is located at $(\rho,z) = (\rho_{12},h)$.
This setup can be modelled by 2 vertical line charge
distributions and their image. Since the emitters are identical in every other respect, they
possess identical line charge density $\Lambda(z)$ of extent $L$. We shall assume that the line charge
density is linear: $\Lambda(z) = \lambda z$. This puts a restriction on the shape of each emitter-base but otherwise
does not pose any limitation on the main conclusions regarding shielding. Thus, in view of the linearity assumption,
we are considering two ellipsoid-like emitters placed a distance $\rho_{12}$ apart.  The potential
at any point ($\rho,z$) can be expressed as \cite{db_fef}

\be
\begin{split}
V(\rho,z) = & \frac{1}{4\pi\epsilon_0}\Big[ \int_{-L}^L \frac{\lambda s}{\big[\rho^2 + (z - s)^2\big]^{1/2}} ds ~
  + \\
  &  \int_{-L}^L \frac{\lambda s}{\big[(\rho - \rho_{12})^2 + (z - s)^2\big]^{1/2}} ds \Big] + E_0 z \label{eq:pot}
\end{split}
\ee

\noi
where the integration is along the $z-$axis from $[-L,L]$, $L$ being
the extent of the line charge distribution along the $z-$axis and $E_0$ is the magnitude of
the asymptotic field or the external field in the absence of the ellipsoidal protrusion.
The parameter $\lambda$ can be fixed by demanding that potential vanishes at the apex. Thus, at
the apex of either emitter,

\be
\begin{split}
  \frac{\lambda}{4\pi\epsilon_0}\Big[ & \int_{-L}^L \frac{s}{\big[(\rho_{12})^2 + (h - s)^2\big]^{1/2}} ds~ + \\
  & \int_{-L}^L \frac{s}{(h - s)} ds   \Big]~ +~ E_0 h = 0. \label{eq:lam}
\end{split}
\ee

\noi
When the two emitters are well separated, the zero-potential contour of the above potential
defines 2 ellipsoidal emitters, each with base radius $b = (h^2 - L^2)^{1/2}$  and separated by $\rho_{12}$, mounted
on a flat planar surface. As the emitters are brought closer, the zero potential contour
keeps the apex invariant due to the imposition of Eq.~\ref{eq:lam}, but its shape 
deviates slightly from an ellipsoid as it approaches the base. The effect gets especially
marked when the separation is small ($\rho_{12}/h < 0.2$) and the linear line charge density
can no longer be used to model the 2-emitter ellipsoidal system. We shall therefore steer clear of this regime.
Furthermore, we shall assume that the deviation in the zero-potential contour of individual emitters away from the
apex, introduces a change in the apex field enhancement factor 
that is small compared to the direct effect of neighbouring emitters at the apex.
Note that for an isolated emitter, the parameter $L = \sqrt{h(h - R_a)}$ \cite{db_fef}. Since the imposition of
Eq.~\ref{eq:lam} preserves the height and apex radius of curvature of the zero-potential surface, this
quantity remains invariant for $\rho_{12}/h > 0.2$.

We are interested here in the field enhancement factor, $\gamma_a$. For axially
symmetric emitters aligned along $\hat{z}$, this is defined as
$\gamma_a = - \frac{1}{E_0} \frac{\partial V}{\partial z} {|_{\rho=0,z=h}}$.
Our starting point for the AFEF is Eq.~\ref{eq:pot}. At the apex, $(\rho,z) = (0,h)$ of emitter 1,

\be
\begin{split}
  \frac{\partial V}{\partial z} {|_{(\rho=0,z=h)}} = & - \frac{\lambda}{4\pi\epsilon_0}\Big[\int_{-L}^L \frac{s}{(h-s)^2} ds~ + \\
    &  \int_{-L}^L \frac{s(h-s)}{\big[\rho_{12}^2 + (h - s)^2\big]^{3/2}} ds   \Big] + E_0
\end{split}
\ee

\noi
which, on integrating, leads to

\be
\begin{split}
  \frac{\partial V}{\partial z} {|_{(\rho=0,z=h)}} = & - \frac{\lambda}{4\pi\epsilon_0}\Big[ \frac{2hL}{h^2 - L^2} +
    \ln\Big(\frac{h+L}{h-L}\Big)  \\
    & +  \int_{-L}^L \frac{s(h-s)}{\big[\rho_{12}^2 + (h - s)^2\big]^{3/2}} ds   \Big] + E_0. \label{eq:Vz0}
\end{split}
\ee

\noi
For $\rho_{12}$ large compared to the base radius of the emitters, the integral in Eq.~\ref{eq:Vz0} is negligible
compared to the first two terms in the square bracket. Furthermore, for sharp emitters ($h/R_a >> 1$), only
the first term dominates \cite{db_fef}. Thus

\be
  \frac{\partial V}{\partial z} {|_{(\rho=0,z=h)}} = - \frac{\lambda}{4\pi\epsilon_0}\Big[ \frac{2hL}{h^2 - L^2} \Big].
    \label{eq:Vz}
\ee

\noi
It now remains to determine $\lambda$ using Eq.~\ref{eq:lam}. The integrals in Eq.~\ref{eq:lam} yield

\be
\begin{split}
  \frac{\lambda}{4\pi\epsilon_0}\Big[& -\int_{-L}^L \frac{h - s}{\big[(\rho_{12})^2 + (h - s)^2\big]^{1/2}} ds~ + \\
    & h \int_{-L}^L \frac{ds}{\big[(\rho_{12})^2 + (h - s)^2\big]^{1/2}} + \\ & h \ln\Big(\frac{h+L}{h-L}\Big) - 
    2L \Big] + E_0 h = 0
  \label{eq:lam1}
\end{split}
\ee

\noi
which further simplifies as

\be
\begin{split}
  \frac{\lambda}{4\pi\epsilon_0}\Bigg[ & \Big[\sqrt{\rho_{12}^2 + (h-s)^2}~ + \\
    & h \ln\Big|\sqrt{1 + \frac{(h+s)^2}{\rho_{12}^2}} + \frac{(h+s)}{\rho_{12}}\Big|~ \Big]_{-L}^L + \\ & h \ln\Big(\frac{h+L}{h-L}\Big) - 
    2L \Bigg] + E_0 h = 0.
  \label{eq:lam2}
\end{split}
\ee

\noi
Note that

\be
\sqrt{\rho_{12}^2 + (h-s)^2}\Big|_{-L}^L \simeq \rho_{12}\Big[1 - \big(1 + 4\delta_{12}^2 \big)^{1/2} \Big]
\ee

\noi
where $\delta_{12} = h/\rho_{12}$. Also,

\be
\ln\Bigg|\sqrt{1 + \frac{(h+s)^2}{\rho_{12}^2}} + \frac{(h+s)}{\rho_{12}}\Bigg|_{-L}^L \simeq \ln\Big|\sqrt{1 + 4\delta_{12}^2} + 2\delta_{12} \Big|
\ee

\noi
if $\rho_{12} >> b$. Finally, using $h \simeq L$ for nano-tipped emitters, we have

\be
\lambda = - \frac{4\pi\epsilon_0 E_0}{\ln\frac{4h^2}{h^2 - L^2} - 2 + \alpha_{S_{12}}} 
\ee

\noi
where the shielding term

\be
\alpha_{S_{12}} = \frac{1}{\delta_{12}}\Big[1 - \sqrt{1 + 4\delta_{12}^2} \Big] + \ln\Big|\sqrt{1 + 4\delta_{12}^2} + 2\delta_{12} \Big|
\ee

\noi
Thus

\be
\gamma_a = \frac{2h/R_a}{\ln\big(4h/R_a\big) - 2 + \alpha_{S_{12}}}
\ee

\noi
is the enhancement factor of the 2-emitter system.

A topic of recent interest has been the change in 2-emitter AFEF as compared to the single or
un-shielded case ($\gamma_a^{(1)}$). Note that for
the isolated or single  case, $\alpha_{S_{12}} = 0$ so that the relative change 

\be
\frac{\gamma_a - \gamma_a^{(1)}}{\gamma_a^{(1)}} = - \frac{\alpha_{S_{12}}}{\Delta} 
\ee

\noi
where $\Delta = \ln\big(4h/R_a\big) - 2$.
For large separations ($\rho_{12}$ large), $\delta_{12}$ is small and it is easy to verify that the

\be
\alpha_{S_{12}} = \frac{2}{3}\delta_{12}^3 + {\cal{O}}(\delta_{12}^5).
\ee

\noi
Thus, at large separations,

\be
\frac{\gamma_a - \gamma_a^{(1)}}{\gamma_a^{(1)}} \sim \frac{1}{\rho_{12}^3}
\ee

\noi
as deduced by other methods \cite{forbes2018,agnol2018}.

\subsection{The N-emitter case}

The 2-emitter case was simple in that everything else being equal, both emitters have the same line charge density.
As a consequence, the shielding term $\alpha_{S_{12}}$ does not depend on $\lambda_{1,2}$.
For a N-emitter system, $N \geq 3$, the line charge densities are unequal since
different emitters have different degrees of shielding. The mutual shielding term
thus generalizes as

\bea
\tilde{\alpha}_{S_{ij}} & = &  \frac{\lambda_j}{\lambda_i} \alpha_{S_{ij}}, ~ \text{where} \\
\alpha_{S_{ij}} & = & \frac{1}{\delta_{12}}\Big[1 - \sqrt{1 + 4\delta_{ij}^2} \Big] + \ln\Big|\sqrt{1 + 4\delta_{ij}^2} + 2\delta_{ij} \Big| \nonumber
\eea

\noi
and $\lambda_j/\lambda_i \neq 1$ in general. The net shielding due to N-emitters
can be expressed as

\be
\tilde{\alpha}_{S_i} = \sum_{j\neq i} {\tilde{\alpha}}_{S_{ij}}.
\ee

\noi
This follows on noting  that Eq.~\ref{eq:Vz} can be expressed for the $i^{th}$ emitter apex located
at ($\rho_{i},h$) with respect to an arbitrary origin, as

\be
  \frac{\partial V}{\partial z} {|_{(\rho=\rho_i,z=h)}} \simeq - \frac{\lambda_i}{4\pi\epsilon_0}\Big[ \frac{2hL}{h^2 - L^2} \Big]     \label{eq:Vzi}
\ee

\noi
since the other terms are small and can be neglected as before. Further, Eq.~\ref{eq:lam}
can be expressed as

\be
\begin{split}
  \frac{\lambda_i}{4\pi\epsilon_0}\Big[ & \sum_{j \neq i} \frac{\lambda_j}{\lambda_i} \int_{-L}^L \frac{s}{\big[(\rho_{12})^2 + (h - s)^2\big]^{1/2}} ds~ + \\
  & \int_{-L}^L \frac{s}{(h - s)} ds   \Big]~ +~ E_0 h = 0 \label{eq:lam3}
\end{split}
\ee

\noi
so that \{$\lambda_i$\} can be determined by simultaneously solving the set of $N$ equations.
We shall instead merely express $\lambda_i$ in terms of the ratio $\lambda_j/\lambda_i$
and use it in Eq.~\ref{eq:Vzi} to express the field enhancement factor

\bea
\gamma_a^{(i)} & = & \frac{2h/R_a}{\ln\big(4h/R_a\big) - 2 + \sum_{j\neq i} \frac{\lambda_j}{\lambda_i} \alpha_{S_{ij}}} \\
& = & \frac{2h/R_a}{\ln\big(4h/R_a\big) - 2 + \tilde{\alpha}_{S_i}}  \label{eq:gamN}
\eea

\noi
for the $i^{th}$ emitter.

Eq.~\ref{eq:gamN} as such does not help us in computing $\gamma_a^{(i)}$
without actually solving the electrostatic problem. However, there are clearly two aspects in the
shielding term that must draw our attention. The first is a geometric factor $\alpha_{S_{ij}}$
which merely depends on the ratio of the height $h$ and the mutual distance $\rho_{ij}$
and does not require a solution of the Poisson equation. The second is the ratio $\lambda_j/\lambda_i$
which does require knowledge about the line charge density. As a first approximation, for
average separations $\rho_{ij}$ comparable to or larger than the height $h$, we
set the ratio $\lambda_j/\lambda_i = 1$. Thus $\tilde{\alpha}_{S_i} = \alpha_{S_i}$ so  that

\bea
\gamma_a^{(i)} & \simeq & \frac{2h/R_a}{\ln\big(4h/R_a\big) - 2 + \sum_{j\neq i} \alpha_{S_{ij}}} \label{eq:gamN0} \\
& = & \frac{2h/R_a}{\ln\big(4h/R_a\big) - 2 + \alpha}_{S_i}.  \label{eq:gamN1}
\eea

\noi

\begin{figure}[htb]
\hspace*{-1.0cm}\includegraphics[width=0.55\textwidth]{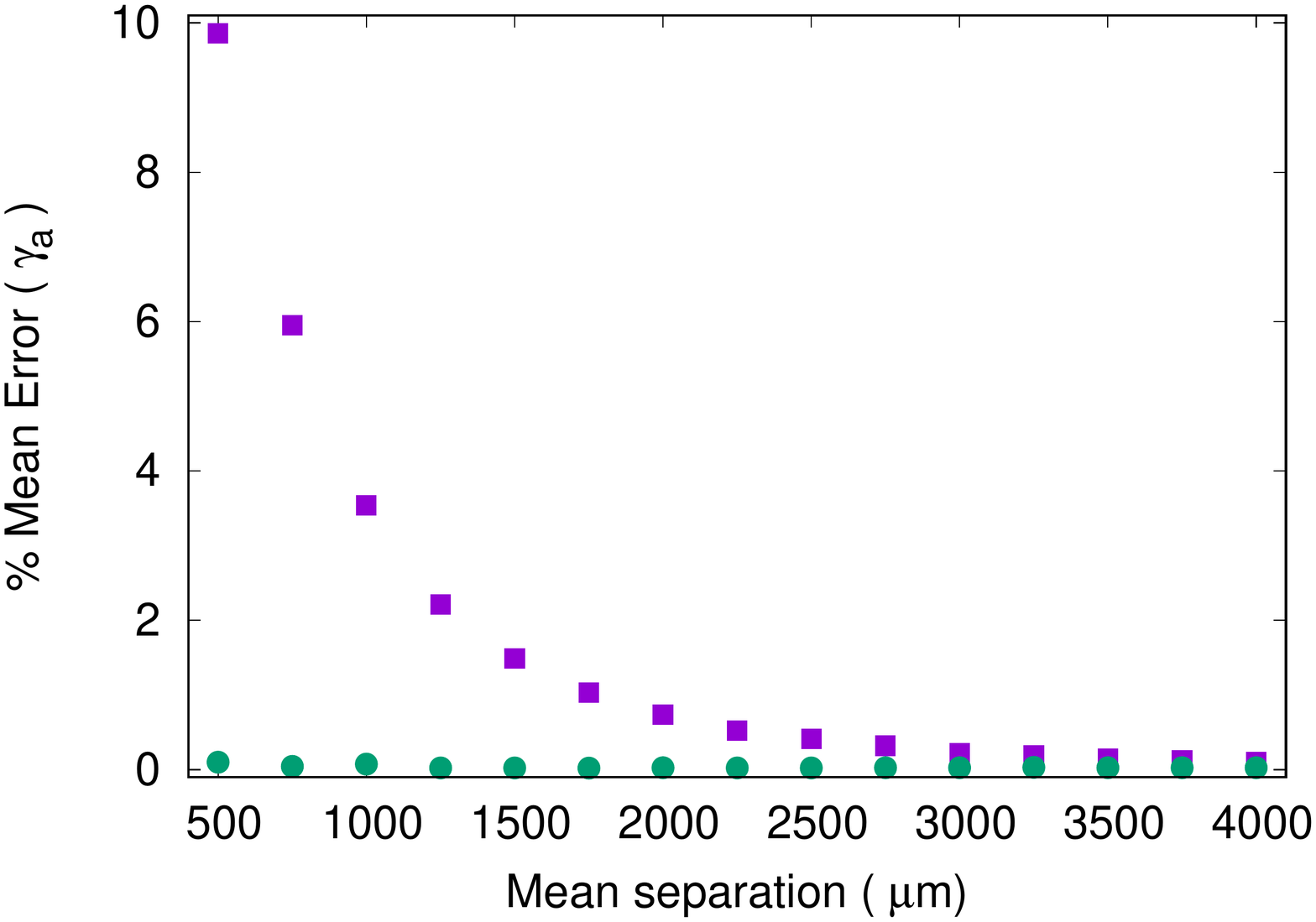}
\caption{The mean error in apex field enhancement factor (AFEF) is shown for different mean spacings.
  A total of 2500 emitters is considered in each case. The field enhancement factor is computed using
  Eq.~\ref{eq:gamN} (denoted by $\CIRCLE$), and, Eq.~\ref{eq:gamN1} (denoted by $\blacksquare$).
  The error in the second case decreases as the mean spacing exceeds the height $h = 1500~\mu m$.
  The emitters have a base radius $b = 10~\mu$m and an apex radius $R_a = 66.67~$nm.
  For Eq.~\ref{eq:gamN}, the error is uniformly small ($\leq 0.05\%$).}
\label{fig:FEF_compare1}
\end{figure}

A comparison of the discrepancy between the AFEF values computed using Eq.~\ref{eq:gamN}
and \ref{eq:gamN1} is shown in Fig.~\ref{fig:FEF_compare1} for a  randomly distributed N-emitter
system. The emitters positions are drawn from a uniform distribution using a standard
random number generator.
It is apparent that the error on ignoring the variation in $\{\lambda_i\}$
is acceptable when the mean separation exceeds the emitter height. Moreover even when the separation
is half the emitter height, the average error is only $6 \%$. Thus, the shielding process
is pre-dominantly a geometric effect and the field enhancement factor can be computed
quite accurately only from a knowledge of the positions, height and apex radius of curvature
of the emitters. We have also determined the mean error of only those emitters for which the AFEF
exceeds the mean value of the AFEF since these predominantly contribute to the field-emission
current at low to medium local field strengths ($< 7$V/nm). The mean error for both Eq.~\ref{eq:gamN} and
Eq.~\ref{eq:gamN1} is then as small as 0.02\% at all spacings considered. Thus,
Eq.~\ref{eq:gamN1} can be used to accurately determine the field enhancement factor
in a uniform random distribution of field-emitters.

\section{The field-enhancement-factor distribution for randomly placed emitters}

As seen above, the field enhancement factor can be computed for individual emitters in a LAFE
purely from geometrical considerations. While this information is useful, a distribution of
field enhancement factors is desirable so that the total current emitted from a LAFE
can be computed based on only a few parameters such as the mean and standard
deviation of the AFEF distribution. We shall thus explore the existence of a universal AFEF
distribution when the emitters are distributed uniformly on a rectangular patch.

\begin{figure}[htb]
\hspace*{-1.0cm}\includegraphics[width=0.55\textwidth]{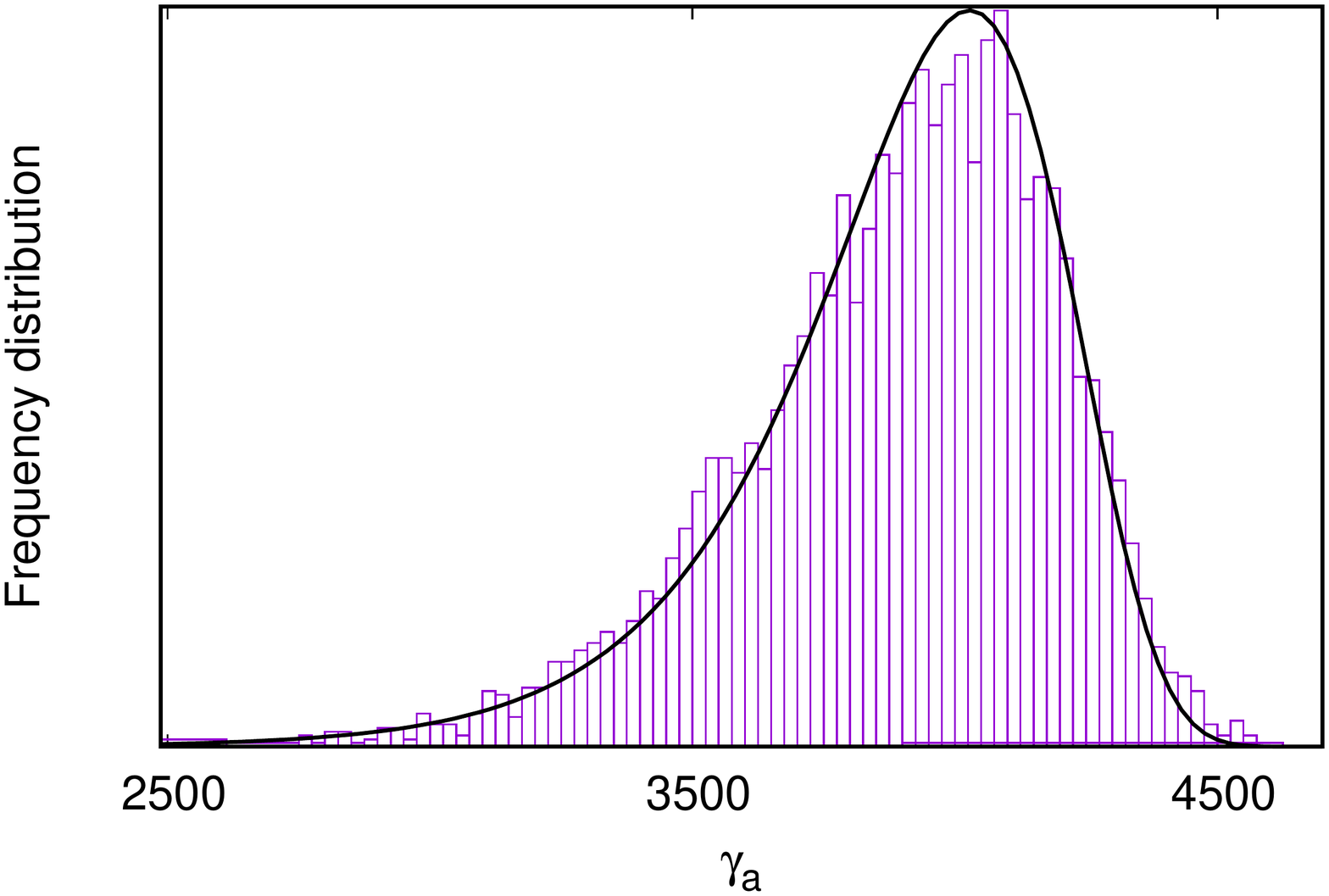}
\caption{The normalized frequency distribution of the field enhancement factors together
  with the Gumbel (minimum) distribution (solid curve). The LAFE has a mean separation $2000~\mu$m
  containing 5000 emitters, each of height  $h = 1500~\mu$ m. The parameters of the Gumbel
  distribution are fixed using Eqns.~\ref{eq:mean} and \ref{eq:sdv} with the mean $\mu$ and
  standard deviation $\sigma$ calculated using the exact numerical values of AFEF which are
  obtained by numerical differentiation of the potential.}
\label{fig:hist}
\end{figure}

Our studies on the field-enhancement-factor distribution for various mean separations, height
and apex radius show that the distribution is closer to a Gaussian when the mean separation
is equal to or somewhat smaller than the emitter height. However, as the mean separation increases,
the field-enhancement-factor distribution is skewed to the left for emitters distributed
uniformly on a rectangular patch as seen in Fig.~\ref{fig:hist}. The skewness persists for
separations beyond twice the emitter height. Also, its mean and standard deviation depend on the
mean separation of emitters. While the mean AFEF increases with separation, the standard
deviation decreases.

An analytical expression for the AFEF distribution is difficult to
derive but fits to various left-skewed distributions show that the Gumbel minimum distribution
best describes the field enhancement factor for mean separations exceeding the emitter
height. This is also the region of interest since as the optimal separation at which the
current density is highest lies here.

The Gumbel distribution has a probability density function

\be
f_\gamma(x) = \frac{1}{\beta} e^{\frac{x-\alpha}{\beta}} e^{-e^{\frac{x-\alpha}{\beta}}}
\ee

\noi
and its cumulative density function is

\be
F_\gamma(x) = 1 - e^{-e^{\frac{x-\alpha}{\beta}}}.
\ee

\noi
Here $\alpha$ and $\beta$ are parameters in terms of which, the mean $\mu$ and
standard deviation $\sigma$ are

\bea
\mu & = & \alpha - \beta \gamma_{EM} \label{eq:mean}\\
\sigma & = & \pi \beta/\sqrt{6} \label{eq:sdv}
\eea

\noi
where $\gamma_{EM} \simeq 0.5772$  is the Euler-Mascheroni constant. Using the numerically
calculated values of apex field enhancement factor $\gamma_a$, $\mu$ and $\sigma$ can be determined. The
Gumbel parameters $\alpha$ and $\beta$ can thus be evaluated using Eqns.~\ref{eq:mean} and \ref{eq:sdv}.
A comparison of the normalized frequency distribution with the Gumbel distribution is shown in Fig.~\ref{fig:hist}.
The agreement is good for mean separations exceeding the emitter height.

\begin{figure}[htb]
\hspace*{-1.0cm}\includegraphics[width=0.575\textwidth]{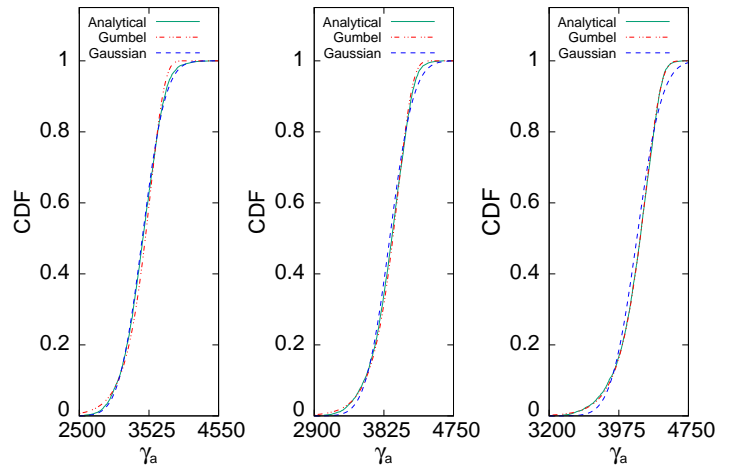}
\caption{A comparison of the cumulative density function (CDF)
  with the Gaussian and Gumbel minimum distributions
  for 3 values of mean separation. The values of $\gamma_a$ are obtained using the
  analytical expression Eq.~\ref{eq:gamN1}. The mean separations are $1500~\mu$m (left),
  $2000~\mu$m (middle) and $2500~\mu$m (right) while the height of the emitters is $1500~\mu$m.
  The parameters for the Gaussian and Gumbel distributions are obtained using the mean
and standard deviation.}
\label{fig:cdf}
\end{figure}

The Gumbel distribution thus shows good agreement when the AFEF values are determined by numerical
differentiation after solving for the electrostatic potential (Eq.~\ref{eq:gamN} 
may instead be used since the errors are small at all separations but requires knowledge
of the line charge density).
We can alternately study the AFEF distribution when
individual AFEF values are determined using Eq.~\ref{eq:gamN1} which does not require
knowledge of the electrostatic problem. The Gumbel distribution again provides a good
approximation when the mean separation between emitters exceeds the emitter height (see Fig.~\ref{fig:cdf})
while the Gaussian distribution is a better approximation for mean separation around the emitter height.
These conclusions hold for other height and apex radius combinations.

\subsection{The harmonic mean and standard deviation using the pair-wise distance distribution}

So far, we have determined the Gumbel parameters by first computing $\gamma_a$ (using Eq.~\ref{eq:gamN} or
\ref{eq:gamN1}) and
determining $\mu$ and $\sigma$. In principle, it should be possible to determine
the mean and standard deviation using Eq.~\ref{eq:gamN1} (but without evaluating individual
$\gamma_a^{(i)}$) and noting that the
emitters are distributed uniformly. It thus requires knowledge of the probability
density function (PDF) of the distance $\rho_{ij}$ where the $i^{th}$ emitter is fixed
and the other $N-1$ emitters distributed uniformly in a rectangular patch. The PDF
obviously depends on the location of the $i^{th}$ emitter and hence various cases
need to be listed. A simpler and well known PDF is that of the pair-wise distance between
any 2 points $i$ and $j$ located on a rectangular patch. This can be used
to calculate the harmonic mean by noting that Eq.~\ref{eq:gamN0} can be rewritten
as

\be
\frac{2h}{R_a} \frac{1}{\gamma_i} = \ln\Big(\frac{4h}{R_a}\Big) - 2 + \sum_{j\neq i} \alpha_{S_{ij}}
\ee

\noi
so that

\be
\frac{2h}{R_a} \sum_i \frac{1}{\gamma_i} = N \Big[ \ln\Big(\frac{4h}{R_a}\Big) - 2 \Big] + \sum_i \sum_{j\neq i} \alpha_{S_{ij}}.
\ee

\noi
Thus, the harmonic mean, $\mu_h$ is

\bea
 \Big(\frac{1}{N} & \sum\limits_i & \frac{1}{\gamma_i} \Big)^{-1}  =   \frac{\frac{2h}{R_a}}{\ln(\frac{4h}{R_a}) - 2 + \frac{1}{N} \sum\limits_i\sum\limits_{j\neq i} \alpha_{S_{ij}}} \\
& = & \frac{\frac{2h}{R_a}}{\ln(\frac{4h}{R_a}) - 2 + (N-1) \int f_\rho(x) \alpha_S(x)} \label{eq:hmean}
\eea

\noi
where

\be
\alpha_S(x) =  \frac{x}{h}\Big[1 - \sqrt{1 + 4\Big(\frac{h}{x}\Big)^2} \Big] + \ln\Big|\sqrt{1 + 4\Big(\frac{h}{x}\Big)^2} + 2\frac{h}{x} \Big|
\ee

\noi
and the probability density function, $f_\rho(x)$, for the pair-wise distance between any two points
distributed uniformly on a square patch of length $L$ is related to  

\be
f_S(s) =
\begin{cases}
  -4 \frac{\sqrt{s}}{L^3} + \frac{\pi}{L^2} + \frac{s}{L^4} & \text{$0 < s \leq L^2$;} \\
  ~~~~~ \\
  -2\frac{1}{L^2} + \frac{4}{L^2}\sin^{-1}\big(\frac{L}{\sqrt{s}}\big)  + \\
  \frac{4}{L^3} \sqrt{s - L^2} - \frac{\pi}{L^2} - \frac{s}{L^4} & \text{$L^2 < s \leq 2L^2$}
  \end{cases}
\ee

\noi
where $s = x^2$ and $f_\rho(x) = 2x f_S(s)$. Note that $\int  f_\rho(x) dx = 1$ so that
Eq.~\ref{eq:hmean} has the factor $(N-1)$ multiplying the integral.
Thus, the harmonic mean can be evaluated, at least numerically for randomly placed ellipsoidal
emitters on a square  (see Philip \cite{philip} for an expression for $f_S(s)$ when the area is rectangular)
patch.

A similar expression can be derived for

\be
\mathlarger\mu_{h2} = \Big(\frac{1}{N} \sum_i \frac{1}{\gamma_i^2} \Big)^{-1}
\ee

\noi
using the probability density function $f_\rho(x)$ and
hence the `harmonic' standard deviation, defined as

\be
\sigma_h = \Big[ \Big(\frac{1}{N} \sum_i \frac{1}{\gamma_i^2} \Big)^{-1} - \Big(\frac{1}{N} \sum_i \frac{1}{\gamma_i} \Big)^{-2} \Big]^{1/2}  \label{eq:sdvh}
\ee

\noi
can be evaluated. In terms of $f_\rho(x)$,

\be
\mathlarger\mu_{h2} \simeq \frac{\big(\frac{2h}{R_a}\big)^2}{\Delta^2 + 2\frac{\Delta}{N} \sum\limits_i\sum\limits_{i\neq j} \alpha_{S_{ij}} + \frac{1}{N} \sum\limits_i \sum\limits_{j\neq i}\sum\limits_{k\neq i} \alpha_{S_{ij}} \alpha_{S_{ik}}  }
\ee

\noi
where $\Delta = \ln(\frac{4h}{R_a}) - 2$. The summations can be expressed as

\be
2\frac{\Delta}{N} \sum\limits_i\sum\limits_{i\neq j} \alpha_{S_{ij}}   =   2\Delta (N-1)   \int f_\rho(x)\alpha_S(x) dx
\ee

\noi
and

\be
\begin{split}
  \frac{1}{N} \sum\limits_i & \sum\limits_{j\neq i}\sum\limits_{k\neq i}  \alpha_{S_{ij}} \alpha_{S_{ik}}  =   (N-1) \int f_\rho(x) \alpha_S^2(x) dx~ + \\
  & (N-1)(N-2) \Big[\int f_\rho(x) \alpha_S(x) dx\Big]^2
  \end{split}
\ee

\noi
and hence $\mathlarger\mu_{h2}$ and $\sigma_h$ can be evaluated in terms of the pair-wise distance distribution $f_\rho(x)$.

\subsection{Gumbel parameters using the harmonic mean and standard deviation}

It is thus possible to evaluate $\mu_h$ and $\sigma_h$ using the pair-wise distribution.
The Gumbel parameters $\alpha$ and $\beta$ can be determined if $\mu_h$ and $\sigma_h$
can be found for the Gumbel distribution as well.

Noting that $\beta$ is generally small compared to $\alpha$ for the field enhancement distribution,
approximate expressions for $\mu_h$ and $\sigma_h$  can be derived as

\bea
\mu_h  & \simeq &  \alpha - \beta \gamma_{EM} - \frac{\beta^2}{\alpha}\frac{\pi^2}{6} \label{eq:muhg} \\
\sigma_h & \simeq & \pi \beta/\sqrt{6} \label{eq:sigmahg}
\eea

\noi
which can be inverted to yield

\bea
\beta & \simeq & \sqrt{6} \sigma_h/\pi  \label{eq:galpha} \\
\alpha & \simeq & \frac{\big(\mu_h + \beta \gamma_{EM}\big) + \sqrt{\big(\mu_h + \beta \gamma_{EM}\big)^2 + \frac{2}{3}\beta^2\pi^2}}{2} \label{eq:gbeta}
\eea

\noi
with $\mu_h$ evaluated using Eq.~\ref{eq:hmean} and $\sigma_h$ using Eq.~\ref{eq:sdvh}. Thus, in
principle, the Gumbel parameters for $N$ uniformly distributed emitters can be evaluated
using the distance distribution $f_\rho$. Note that the expressions for $\mu_h$ and $\sigma_h$ above
are only approximate and hence the Gumbel parameters computed this way are not expected to be accurate.
However, they can be calculated using
the pair-wise distance distribution alone. Table~I shows a comparison of the
Gumbel parameters for 3 different mean separations and in each case, 3 different methods
are adopted to determine the parameters $\alpha$ and $\beta$.  It is clear
that a reasonably good approximation to the Gumbel distribution parameters can be obtained
using the pairwise distribution function.

\begin{table}[tbh]
  \label{tab:gumbel}
\begin{center}
    \begin{tabular}{|c| c | c | c |}
      \hline
      Separation & Method & Gumbel $\alpha$ & Gumbel $\beta$ \\ \hline \hline
      1500 $\mu$m & numerical with $\mu$ and $\sigma$ & 3548 & 211 \\ \cline{2-4}
      & numerical with $\mu_h$ and $\sigma_h$ & 3543 & 217 \\ \cline{2-4}
      & pair-wise with $\mu_h$ and $\sigma_h$ & 3535 & 211 \\ \hline \hline
      2000 $\mu$m & numerical with $\mu$ and $\sigma$ & 4015 & 195 \\ \cline{2-4}
      & numerical with $\mu_h$ and $\sigma_h$ & 4015 & 205 \\ \cline{2-4}
      & pair-wise with $\mu_h$ and $\sigma_h$ & 4012 & 206 \\ \hline \hline
      2500  $\mu$m & numerical with $\mu$ and $\sigma$ & 4276 & 174 \\ \cline{2-4}
      & numerical with $\mu_h$ and $\sigma_h$ & 4279 & 184   \\ \cline{2-4}
      & pair-wise with $\mu_h$ and $\sigma_h$ & 4277 & 190 \\ \hline
    \end{tabular}
\end{center}

\caption{The Gumbel parameters evaluated using three methods for each separation. For the first two
methods  (`numerical'),  the mean $\mu~ \text{or}~ \mu_h$ and standard deviation $\sigma~ \text{or}~ \sigma_h$
are evaluated using approximate analytical values of $\gamma_a$ obtained using Eq.~\ref{eq:gamN1}
for a given uniform distribution of emitter positions. These are then equated to the corresponding
expressions for the Gumbel minimum distribution in order to determine $\alpha$ and $\beta$. The third method uses
the pairwise distance distribution to evaluate $\mu_h$ and $\sigma_h$ directly using numerical integration.
Note that the second
and third methods use approximate expressions for $\mu_h$ (Eqn.~\ref{eq:muhg}) and $\sigma_h$ (Eq.~\ref{eq:sigmahg}).
}
\end{table}

\section{Current from a LAFE}

In the previous sections, we have derived a formula for the apex field enhancement factor (AFEF) $\gamma_a$
of individual emitters in a LAFE. In addition, we have found that the Gaussian distribution approximates the
AFEF distribution well when the mean separation is close to the emitter height while the Gumbel minimum distribution
is a better approximation, for mean separation larger than the height of
individual emitters. Assuming recent results on the variation of field enhancement factor
around the apex \cite{db_ultram,db2018_pop1}
of individual emitters in a LAFE, the net current emitted can be expressed as

\begin{figure}[t]
\hspace*{-1.0cm}\includegraphics[width=0.5\textwidth]{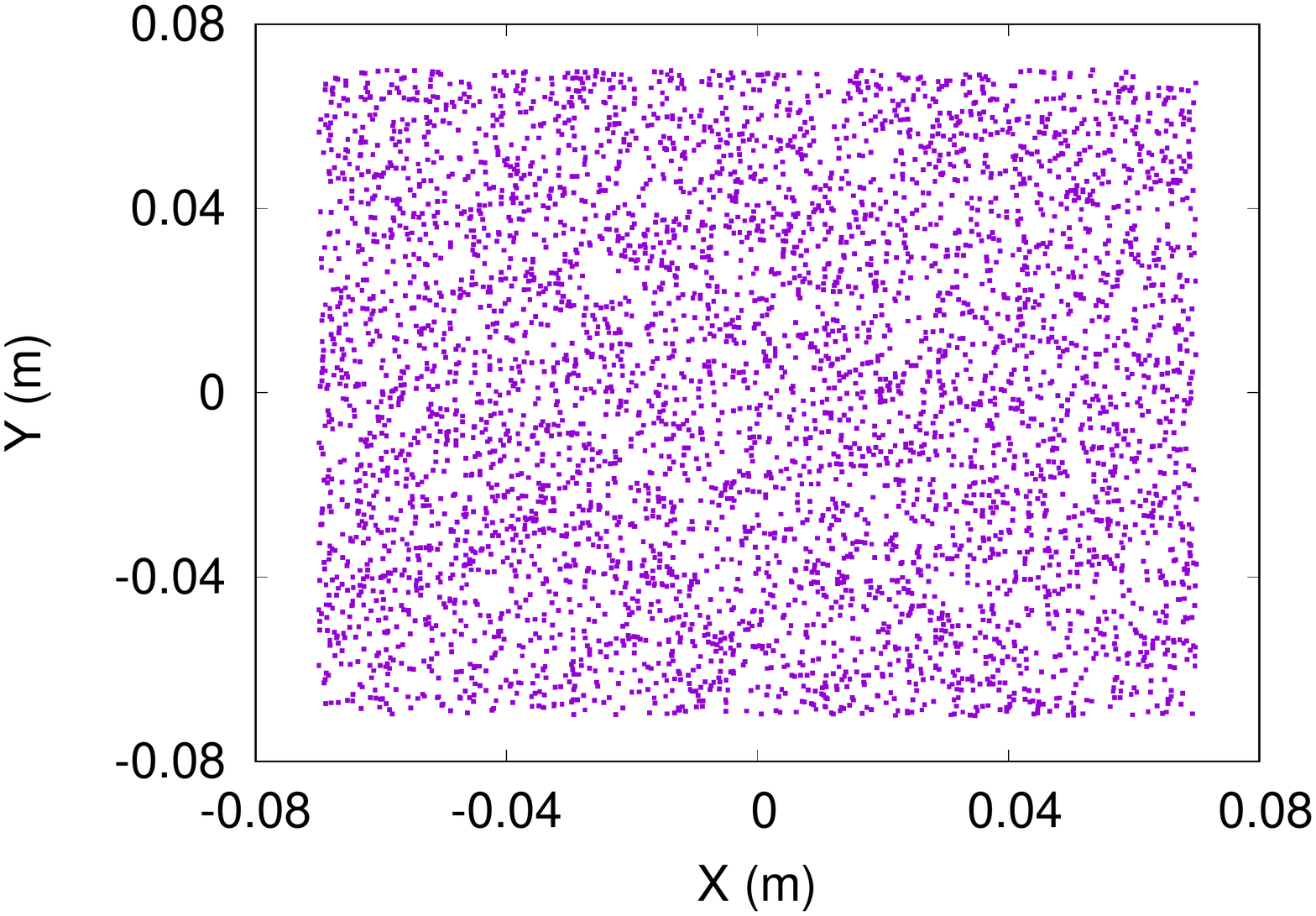}
\caption{A uniformly distributed LAFE. Each point denotes the position of an emitter.}
\label{fig:pos}
\end{figure}

\begin{figure}[h]
\hspace*{-1.0cm}\includegraphics[width=0.6\textwidth]{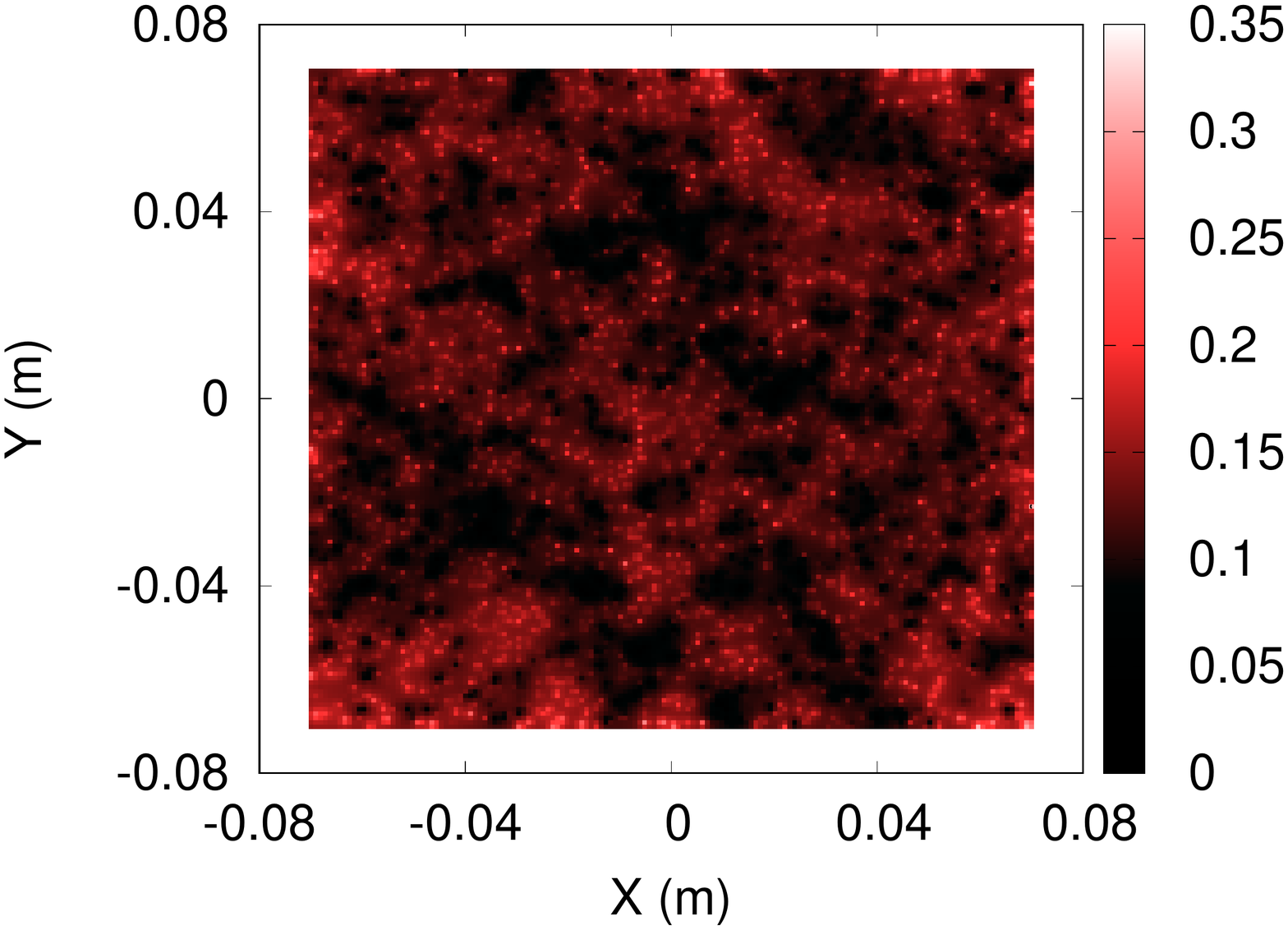}
\caption{The corresponding current map of the LAFE. The dark regions have low electron emission.
The current is measured in amperes.}
\label{fig:cur}
\end{figure}

\be
I_{{\bf LAFE}} = \sum\limits_i I_{i}  = 2 \pi R_a^2 \sum\limits_i J_{a_i} \cg_i \label{eq:totI}
\ee

\noi
where $I_{a_i}$ is the current from the $i^{\rm th}$ emitter and the corresponding apex current density is

\be
J_{a_i} = \frac{1}{t_{F_i}^2} \frac{A_\fn}{\phi} E_{a_i}^2 ~
e^{-B_\fn \mathlarger\nu_{F_i} \phi^{3/2}/E_{a_i}}. \label{eq:FN1}
\ee

\noi
while the area-factor is

\be
\cg_i = \frac{E_{a_i}}{B_{\fn} \phi^{3/2}}  \frac{1}{(1 - f_{0i}/6)}.
\ee

\noi
In the above, $E_{a_i}~=~\gamma_{a_i}~E_0$ is the local field at the apex of the $i^{\rm th}$ emitter, $\gamma_{a_i}$ is the
apex enhancement factor and $E_0$ is the asymptotic electric field.
Here, $A_\fn~\simeq~1.541434~{\rm \mu A~eV~V}^{-2}$ and 
$B_\fn~\simeq~6.830890~{\rm eV}^{-3/2}~{\rm V~nm}^{-1}$ are the conventional FN constants,  $\phi$
is the work function while $\mathlarger\nu_{F_i} \simeq 1 - f_{0i} + \frac{1}{6}f_{0i}\ln f_{0i}$ and $t_{F_i} \simeq$~$1 + f_{0i}/9 - \frac{1}{18}f_{0i}\ln f_{0i}$ are correction factors due to the
image potential with $f_{0i}  \simeq c_S^2 E_{a_i}/\phi^2$ and $c_S^2 = 1.439965~{\rm eV^2~V^{-1}~nm}$.
Unless otherwise stated, the work function $\phi = 4.5$eV in this paper.

Fig.~\ref{fig:pos} shows the position of 4900 uniformly distributed emitters with mean separation $2000~\mu$m.
The corresponding current map is shown in Fig.~\ref{fig:cur}. It can be seen that sparse regions have high
electron emission while denser regions have lower emission.

\begin{figure}[thb]
\vskip -2.1 cm
\hspace*{-1.0cm}\includegraphics[width=0.6\textwidth]{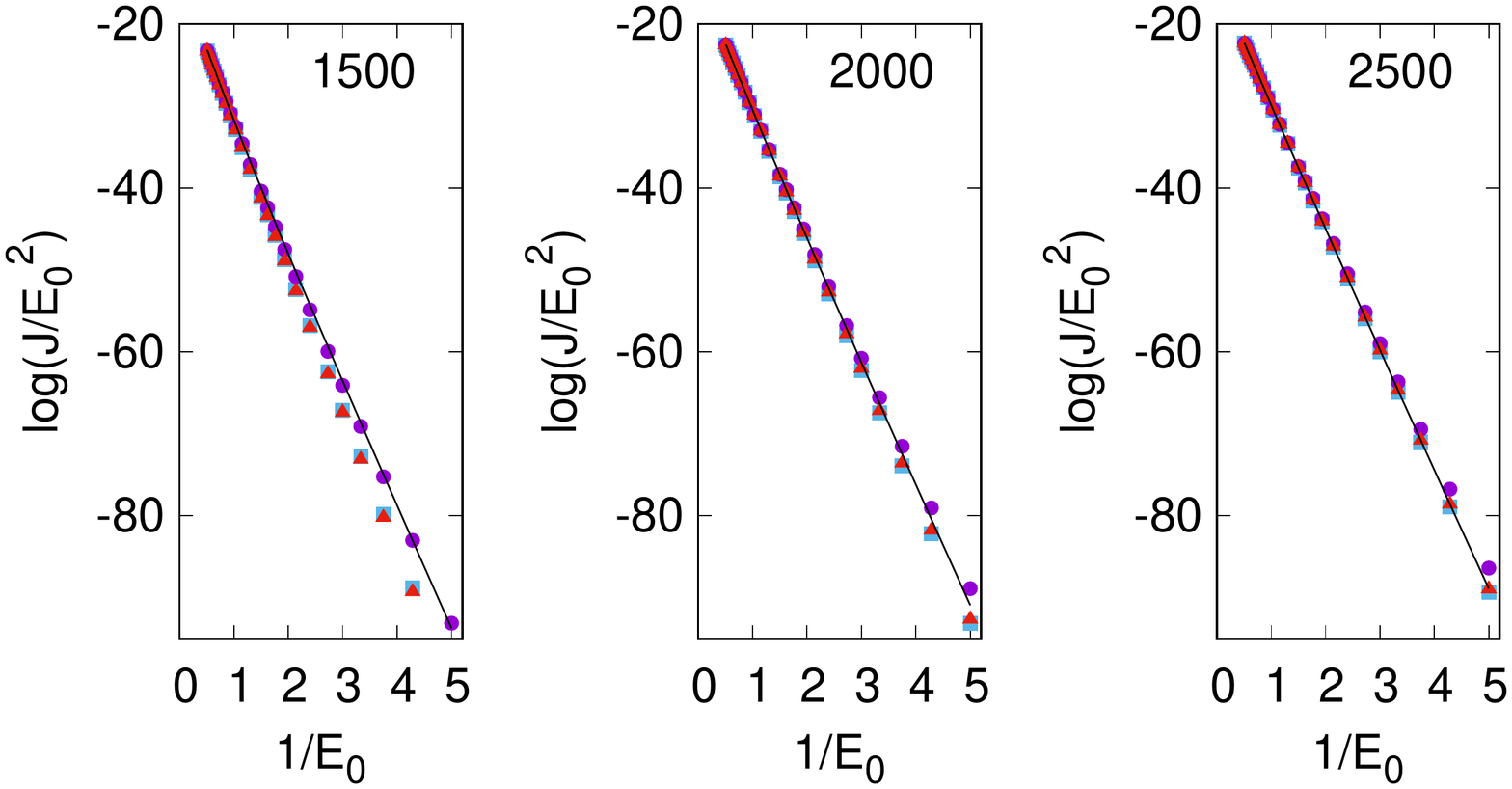}
\vskip -2.4 cm
\hspace*{-1.0cm}\includegraphics[width=0.6\textwidth]{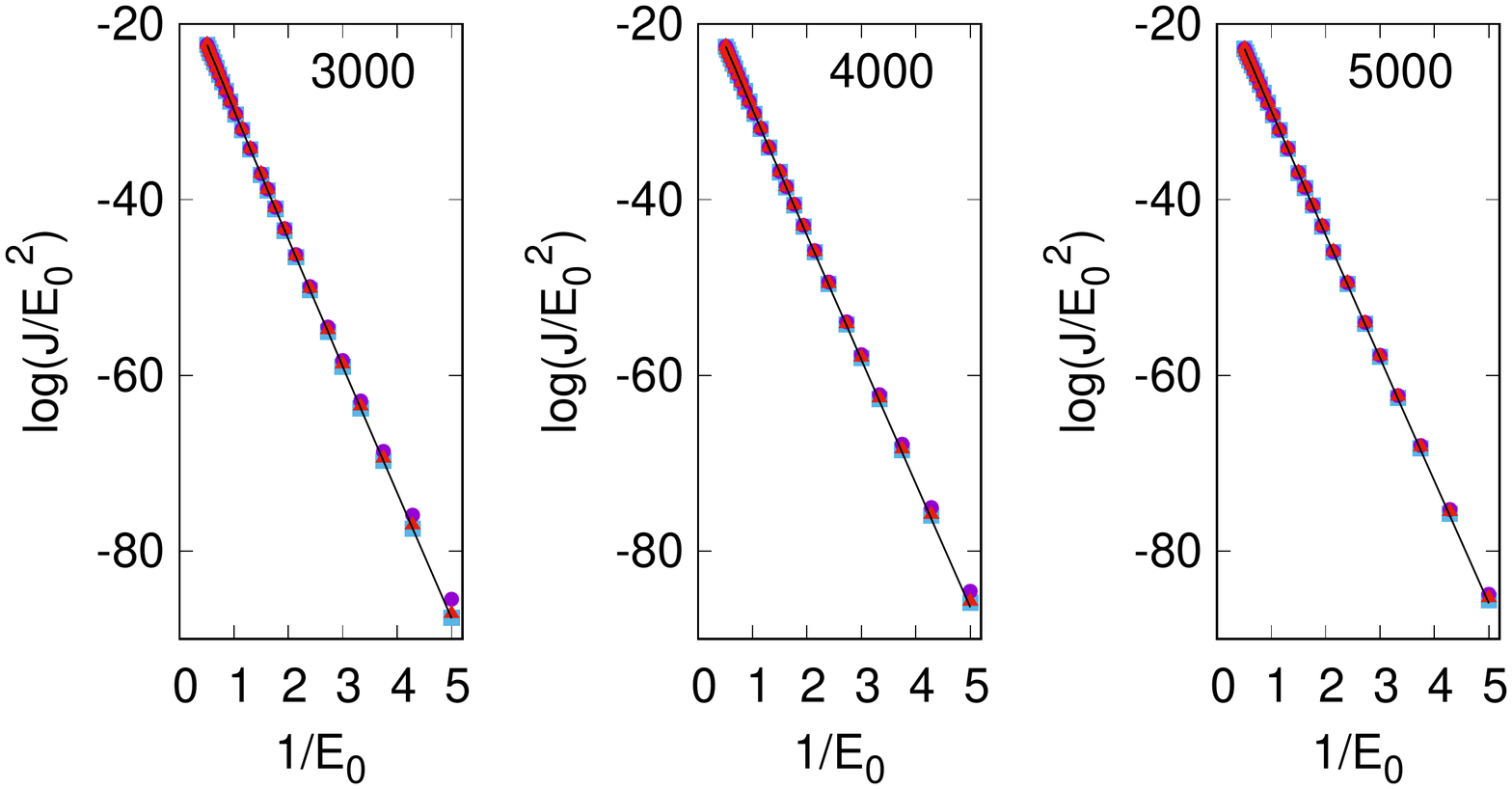}
\vskip -1.0 cm
\caption{A comparison of the FN plots for the LAFE current density for 6 mean separations
found by (a)  summing individual emitter currents using exact AFEF (continuous curve)
(b) using Gumbel distribution and approximate analytical $\gamma_a^{(i)}$ (Eq.~\ref{eq:gamN1})
to find $\mu$ and $\sigma$ and hence the
Gumbel parameters ($\blacksquare$) (c) using Gumbel distribution and
approximate analytical $\gamma_a^{(i)}$ (Eq.~\ref{eq:gamN1})
to find  $\mu_h$ and $\sigma_h$  using pair-wise
distribution ($\blacktriangle$) (d) using Gaussian distribution and approximate analytical
$\gamma_a^{(i)}$ (Eq.~\ref{eq:gamN1}) to find $\mu$ and $\sigma$ ($\CIRCLE$). 
The mean separations are $1500~\mu$m (top-left), $2000~\mu$m (top-middle), $2500~\mu$m (top-right),
$1500~\mu$m (bottom-left), $2000~\mu$m (bottom-middle), $2500~\mu$m (bottom-right).
Cases (b) and (c) are virtually indistinguishable. The X-axis has units of $\mu$m/V, while
$J/E_0^2$ has units of $A/V^2$. The Gaussian distribution over-estimates the current for mean separations
greater than $2000~\mu$m. }
\label{fig:FN1}
\end{figure}

The net current from the LAFE can alternately be calculated using the apex field enhancement
factor distribution. Using an AFEF distribution, the summation in Eq.~\ref{eq:totI}
can be expressed as

\be
I_{{\bf LAFE}} = \sum\limits_i I_{i}  =
2 \pi R_a^2 \int\limits_0^\infty J_a(x) \cg(x) f_\gamma(x) dx \label{eq:totI1}
\ee

\noi
where $f_\gamma(x)$ is the Gumbel distribution and

\bea
J_a(x) &  = &  \frac{1}{t_{F}^2(x)} \frac{A_\fn}{\phi} E_{a}^2(x) ~
e^{-B_\fn \mathlarger\nu_{F}(x) \phi^{3/2}/E_{a}(x)}  \\
\cg(x) & = & \frac{E_{a}(x)}{B_{\fn} \phi^{3/2}}  \frac{1}{(1 - f_0(x)/6)} \\
\mathlarger\nu_{F}(x) & = & 1 - f_0(x) + \frac{1}{6}f_0(x)\ln f_0(x) \\
t_F(x) & \simeq & 1 + \frac{1}{9}f_0(x) - \frac{1}{18}f_0(x)\ln f_0(x) \\
f_0(x) & \simeq & c_S^2 E_{a}(x)/\phi^2 .
\eea

\begin{figure}[thb]
\hspace*{-1.0cm}\includegraphics[width=0.575\textwidth]{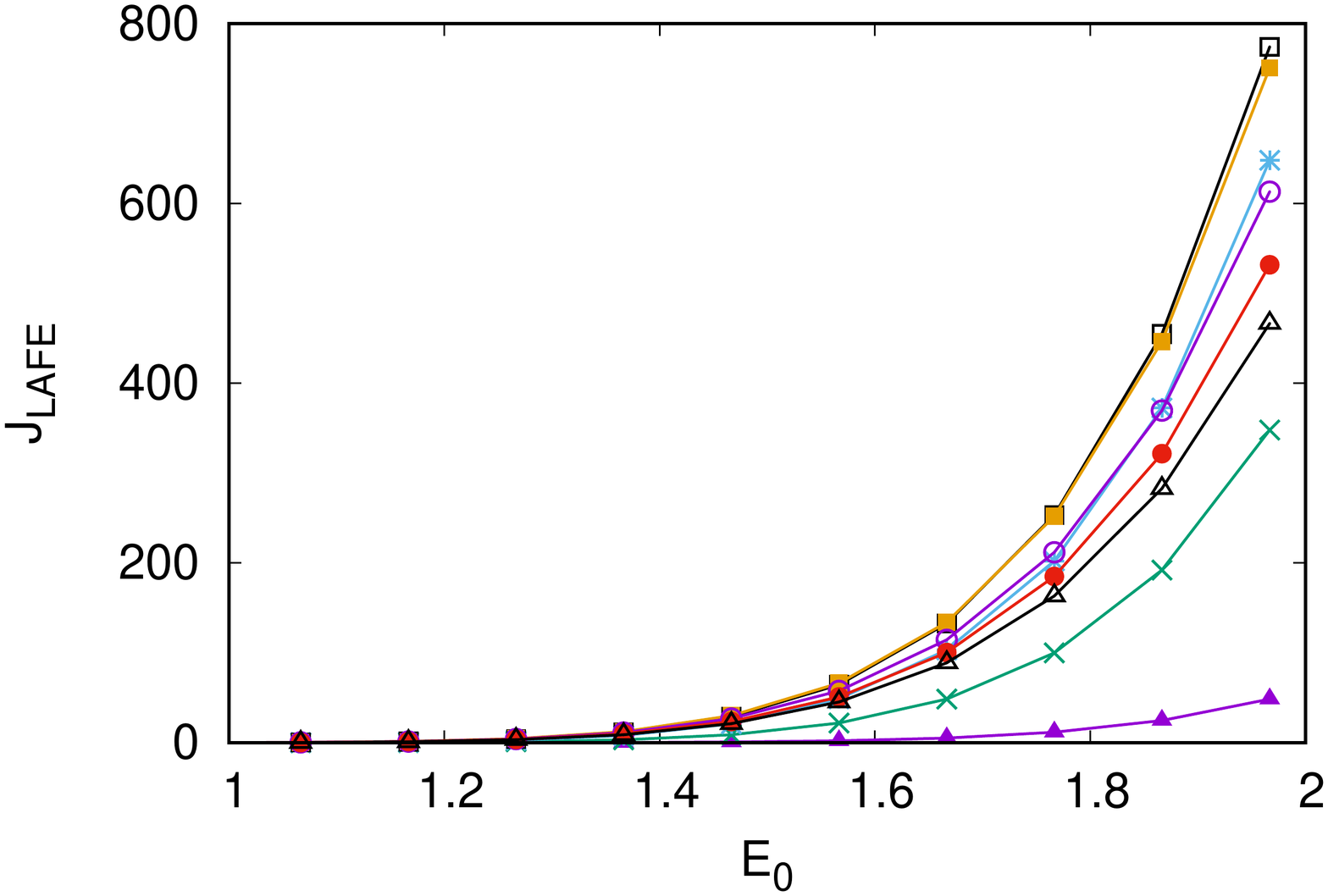}
\vskip -1.0 cm
\caption{A comparison of the exact LAFE current density ($\text{A/m}^2$) for different mean separations
  is shown as a function of the asymptotic field $E_0~\text{(V/}\mu\text{m)}$. The mean separations in units of $\mu$m  are
  1000 ($\blacktriangle$), 1500 (${\boldsymbol{\times}}$), 2000 (${\boldsymbol{\ast}}$),
  2500 (${\boldsymbol{\Box}}$), 3000 ($\blacksquare$), 4000 (${\bigcirc}$), 4500 ($\CIRCLE$) and 5000 (${\boldsymbol{\triangle}}$).
}
\label{fig:JvsE}
\end{figure}

In Fig.~\ref{fig:FN1}, the net current obtained by summing over individual pins (continuous curve) is compared with the
current obtained using Gaussian/Gumbel AFEF distributions,
as a Fowler-Nordheim (FN) plot for different mean separations.  The parameters of the Gumbel
distribution are obtained in 2 ways. In the first ($\blacksquare$),
the mean and standard deviation of $\{\gamma_a^{(i)} \}$ (obtained
from Eq.~\ref{eq:gamN1} for a given realization of uniform distribution),
are used to evaluate $\alpha$ and $\beta$ (Eqns.~\ref{eq:mean} and \ref{eq:sdv}).
In the second ($\blacktriangle$),  $\alpha$ and $\beta$
are obtained using the harmonic mean and standard deviation which in turn are
evaluated using the pairwise distribution. Also shown is the current obtained using a Gaussian
distribution  with $\mu$ and $\sigma$ obtained using $\{\gamma_a^{(i)} \}$ ($\CIRCLE$).
The agreement with the Gumbel distribution is reasonably good at all the separations except when
the mean separation equals the emitter height where the Gaussian distribution performs much better especially at lower
field strengths. The
Gumbel distribution performs well even at low field strengths and notably even when the parameters are
obtained using the pairwise distribution. Note that even though a wide range of external asymptotic
fields $E_0$ has been investigated, for practical purposes, values of $\ln(J/E_0^2)$ exceeding -50
are relevant.

\begin{figure}[thb]
\hspace*{-1.0cm}\includegraphics[width=0.575\textwidth]{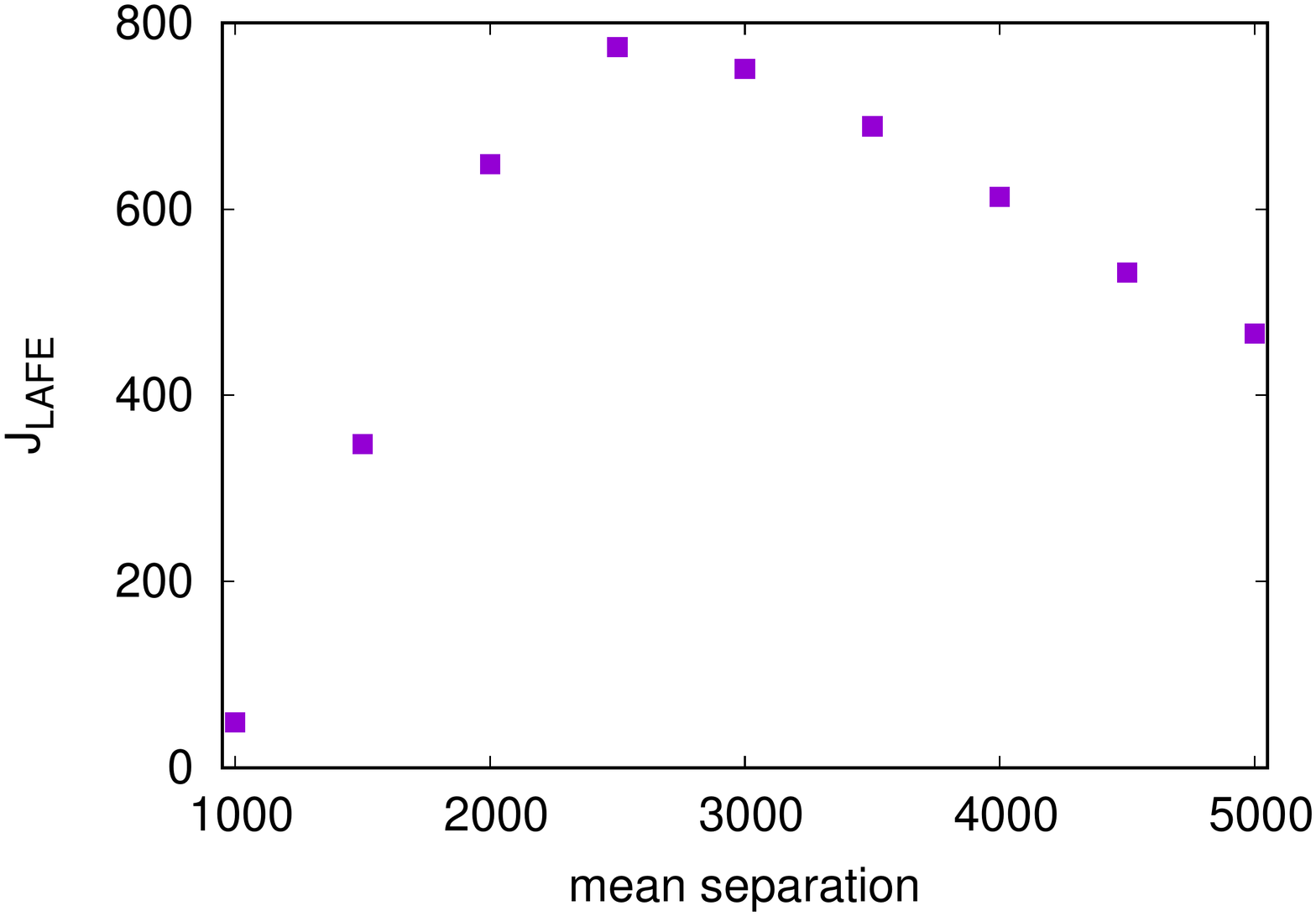}
\vskip -0.10 cm
\caption{Variation of the exact LAFE current density ($\text{A/m}^2$) with mean separation ($\mu$m)
  at $E_0 \simeq 1.96$~V/$\mu$m. 
}
\label{fig:JvsD}
\end{figure}

Finally, we investigate the optimal mean separation at which the emission current
density for the random LAFE considered here is maximum. Fig.~\ref{fig:JvsE} shows
the current density plotted against the asymptotic electrostatic field $E_0$ for 
various mean separations. The maximum current density peaks sharply after the mean separation crosses
the emitter height (1500 $\mu$m) and plateaus at around 2500-3000 $\mu$m
for all field strengths (see Fig.~\ref{fig:JvsD} for the variation of current density
with mean separation at $E_0 \simeq 1.96$~V/nm). This is similar to
the trend seen for an infinite array \cite{jap2016}.

\section{Summary and Conclusions}

We have studied shielding effects in a random large area field emitter starting with a 2-emitter
system. Methods similar to one used in recently \cite{db_fef} were used to first arrive at
a formula for the apex field enhancement factor (AFEF) in a 2-emitter system and this was subsequently
generalized for an arbitrary N-emitter system where the emitter placements may be in a line, a 2-dimensional
array or even randomly distributed. It was found that for purposes of field emission where emitters
with the largest AFEFs contribute, the shielding
effect can be considered to be purely geometric and the AFEFs can be determined, within acceptable limits, purely
from the emitter locations without solving the full electrostatic problem.

The question of AFEF distribution was subsequently investigated. It was found that the distribution
is closer to a Gaussian when the mean separation is close to or somewhat less than the emitter
height, but is better approximated by a Gumbel minimum distribution for spacings larger than
emitter height. It is in this regime that the maximum LAFE current density is found to lie.

These results are supported by computation of the emission current, both, by directly summing over
individual pins after solving the full electrostatic problem, and at the other extreme by using
the expression for approximate (geometric) analytical AFEFs together with the Gumbel and Gaussian distributions.
The comparison shows that the latter method with Gumbel distribution can be used profitably for a large range of field strengths
for mean separations larger than the emitter height but at mean separations close to the emitter height, the
Gaussian distribution performs much better.

Finally, for the uniform distribution of emitters, we have evaluated the Gumbel parameters using the
pair-wise distance distribution by calculating the harmonic mean and standard deviation. The evaluation
of current using these parameters gives excellent results (for mean separation greater than emitter height)
that are virtually indistinguishable from results with parameter obtained from a given realization of
the emitter pins.

\section{Acknowledgements}

The authors thank Raghwendra Kumar, Gaurav Singh and Rajasree for useful discussions.

\vskip 0.05 in
\section{References} 
\vskip -0.25 in

\end{document}